\documentclass[11pt,twocolumn]{article}
\usepackage[T1]{fontenc}
\usepackage{graphicx}
\usepackage{booktabs}
\usepackage{amsmath,amssymb,amsfonts}
\usepackage[margin=1in]{geometry}
\usepackage{authblk}
\usepackage{stfloats}

\begin{document}

\title{\textbf{AMES: Approximate Multi-modal Enterprise Search via Late Interaction Retrieval}}

\author[1]{Tony Joseph}
\author[1]{Carlos Pareja}
\author[1]{David Lopes Pegna}
\author[1]{Abhishek Singh}

\affil[1]{Apple, Cupertino, CA, USA}

\maketitle

\begin{abstract}
We present AMES (Approximate Multimodal Enterprise Search), a unified multimodal late interaction retrieval architecture which is backend agnostic. AMES demonstrates that fine-grained multimodal late interaction retrieval can be deployed within a production grade enterprise search engine without architectural redesign. Text tokens, image patches, and video frames are embedded into a shared representation space using multi-vector encoders, enabling cross-modal retrieval without modality specific retrieval logic. AMES employs a two-stage pipeline: parallel token level ANN search with per document Top-M MaxSim approximation, followed by accelerator optimized Exact MaxSim re-ranking. Experiments on the ViDoRe V3 benchmark show that AMES achieves competitive ranking performance within a scalable, production ready Solr based system.
\end{abstract}


\section{Introduction}

Technical documents in engineering and manufacturing are inherently multimodal, combining text with diagrams, tables, screenshots, and video. In such corpora, relevant evidence is often localized in specific visual regions or temporal segments rather than captured by global document summaries. As a result, lexical retrieval and single vector dense encoders often fail to preserve the fine grained alignment needed between queries and localized content.

Late-interaction multi-vector retrieval addresses this limitation by representing queries and documents as sets of token or patch level embeddings and scoring them with MaxSim style aggregation~\cite{khattab2020colbert}. Although this paradigm has shown strong effectiveness in text and selected multimodal retrieval settings, practical deployment remains limited. Existing systems often depend on specialized ANN infrastructure or custom retrieval engines, and multimodal extensions are rarely evaluated in enterprise search environments.

We present \textbf{AMES} (Approximate Multimodal Enterprise Search), a unified multimodal late-interaction retrieval architecture with a reference implementation on Apache Solr using native vector search and parent/child indexing. AMES stores text tokens, image patches, and video frame embeddings as child documents under shared parent units, preserving metadata filtering, access control, and compatibility with enterprise search workflows. More broadly, the design is backend agnostic and applies to any vector enabled retrieval system supporting ANN search, structured filtering, and parent/child or equivalent grouping semantics.

AMES uses a two stage retrieval pipeline. Stage~1 performs a parallel token-level ANN search over child embeddings and aggregates results client-side using a per-document Top-$M$ MaxSim approximation. When multiple modalities are indexed, the ANN search is performed separately per modality and fused through robust normalization and weighted combination. Stage~2 re-ranks the shortlisted parent documents using Exact MaxSim over all associated child embeddings, implemented with accelerator optimized batched matrix operations in PyTorch. This Approximate $\rightarrow$ Exact design balances scalability and retrieval fidelity without requiring specialized vector retrieval infrastructure.

The same framework extends naturally to video retrieval by segmenting videos into semantically coherent units and indexing frame level embeddings within the same parent/child structure, without changing retrieval logic beyond preprocessing.

To our knowledge, AMES is the first unified multimodal late-interaction retrieval system implemented within a Lucene-based enterprise search engine while preserving fine grained retrieval semantics, metadata aware filtering, and operational compatibility with existing search pipelines.

\subsection{Contributions}

Our main contributions are:
\begin{itemize}
    \item \textbf{AMES}, an enterprise compatible and backend-agnostic architecture for multimodal late-interaction retrieval, demonstrated in Solr;
    \item a parallel token level ANN candidate generation method with per-document Top-$M$ MaxSim aggregation;
    \item modality aware Stage~1 fusion through robust normalization and weighted combination of text and image evidence;
    \item accelerator optimized Exact MaxSim reranking; and
    \item a unified extension to video retrieval without modality specific redesign.
\end{itemize}

Together, these results show that fine-grained multimodal late-interaction retrieval can be deployed within standard enterprise search infrastructure, narrowing the gap between recent multi-vector retrieval advances and production search systems.

\section{Related Work}

\subsection{Late-Interaction Retrieval}

Late-interaction retrieval addresses the limitations of single vector dense representations by preserving token level structure during matching. ColBERT~\cite{khattab2020colbert} represents queries and documents as sets of contextualized token embeddings and scores them with MaxSim aggregation, enabling localized matching and improved effectiveness over bi-encoder models. Subsequent systems such as ColBERTv2~\cite{santhanam2022colbertv2} and PLAID further improve efficiency through compression, indexing, and pruning strategies, showing that late interaction can scale in practice. However, these approaches typically assume specialized ANN infrastructure or custom retrieval engines. In contrast, AMES shows that multi-vector late-interaction can be implemented directly within a production grade enterprise search engine.

\subsection{Multimodal Late-Interaction Retrieval}

Late interaction has been extended beyond text to multimodal retrieval. ColPali~\cite{faysse2024colpali} applies ColBERT-style matching to document images using visual patches, enabling retrieval over visually rich pages containing text, tables, and figures. FLMR~\cite{lin2023flmr} and CLaMR~\cite{wan2025clamr} study fine-grained text-image retrieval for multimodal QA and knowledge grounding, while Video-ColBERT~\cite{reddy2025videocolbert} extends MaxSim-style scoring to frame or region level video representations. These works demonstrate that late-interaction generalizes naturally across modalities, but is typically evaluated in task-specific or specialized retrieval settings. Instead, AMES targets unified retrieval across text, images, and video within a single index and a common retrieval pipeline.

\subsection{Enterprise Search and Multimodal Retrieval}

Recent multimodal encoders such as ColQwen2.5 and ColQwen2.5-Omni embed text, images, and video into shared multi-vector spaces, enabling fine grained cross-modal matching. In parallel, enterprise search engines such as Solr and Elasticsearch have added native vector search to Lucene based infrastructures, alongside support for lexical retrieval, filtering, and access control. However, these systems are typically built around single vector representations and do not natively support fine grained late-interaction retrieval.

Our work connects these directions by combining multimodal multi-vector encoders with a Lucene based enterprise search engine. AMES uses Solr's native vector search, parent/child indexing, and filtering to support token, patch, and frame level retrieval while preserving enterprise metadata semantics and operational compatibility. For evaluation, we use the ViDoRe V3 benchmark~\cite{vidore2026}, whose industrial subset most closely matches the multimodal enterprise retrieval setting considered in this work.

\section{System Overview}

AMES is a multimodal late-interaction retrieval system for heterogeneous technical content. It follows an end-to-end pipeline of segmentation, embedding, indexing, approximate candidate generation, and exact re-ranking. The architecture separates embedding computation from search infrastructure, allowing deployment on vector enabled backends that support ANN retrieval and structured filtering. 

\begin{figure*}[t]
\centering
\includegraphics[width=0.9\textwidth]{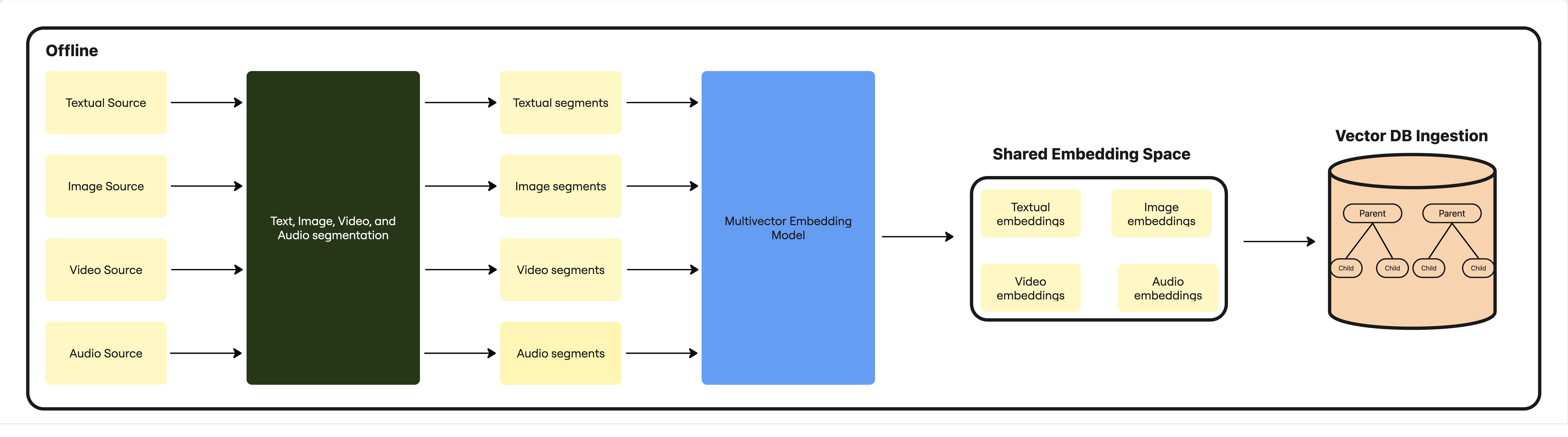}
\caption{Offline indexing pipeline. Documents and media are segmented into retrieval units, encoded with a multi-vector model, and indexed using a parent--child or equivalent grouping schema for ANN candidate generation and Exact MaxSim reranking.}
\label{fig:system_overview}
\end{figure*}

\subsection{Problem Definition and Notation}

Multimodal late-interaction retrieval over enterprise corpora can be formalized as follows.

\subsubsection{Corpus Structure}

Let the corpus consist of a set of parent documents:
\[
\mathcal{D} = \{d_1, d_2, \ldots, d_{|\mathcal{D}|}\}
\]

Each parent document $d$ corresponds to a retrievable unit such as a page, image, or video segment. A parent document represents the atomic retrievable unit returned to the user.

Each parent document is associated with a set of child embeddings:
\[
\mathcal{D}_d = \{\mathbf{d}_j\}_{j=1}^{|\mathcal{D}_d|}
\]
where each $\mathbf{d}_j \in \mathbb{R}^m$ is a dense vector representing a textual token, an image patch, or a video frame region.

All child embeddings are produced by a shared multimodal encoder and lie in a common embedding space $\mathbb{R}^m$. All modalities are embedded using a unified multimodal encoder trained to align textual, visual, and temporal representations in a shared embedding space.

\subsubsection{Query Representation}

Given a textual query $q$, the encoder produces a set of token-level embeddings:
\[
\mathcal{Q} = \{\mathbf{q}_i\}_{i=1}^{|\mathcal{Q}|}
\]
where each $\mathbf{q}_i \in \mathbb{R}^m$.

All embeddings are L2-normalized allowing us to compute similarity using dot product.
\[
\|\mathbf{q}_i\|_2 = 1, \quad \|\mathbf{d}_j\|_2 = 1
\]

\subsubsection{Late Interaction Scoring}

Relevance between query $\mathcal{Q}$ and parent document $d$ is defined using MaxSim aggregation:
\[
\text{MaxSim}(\mathcal{Q}, d) = \sum_{i=1}^{|\mathcal{Q}|} \max_{\mathbf{d}_j \in \mathcal{D}_d} \langle \mathbf{q}_i, \mathbf{d}_j \rangle
\]

This token/region-level aggregation defines the late-interaction relevance score.

\subsubsection{Two-stage retrieval objective}

Because computing exact MaxSim over the full corpus is computationally expensive, we adopt a two-stage retrieval objective:

First we perform an approximate candidate generation producing a candidate set
\[
\mathcal{R}(\mathcal{Q}) \subset \mathcal{D}, \quad |\mathcal{R}(\mathcal{Q})| = N
\]

Then we perform exact re-ranking computing the full MaxSim only over $\mathcal{R}(\mathcal{Q})$.

Exact MaxSim for a single parent document requires $O(|\mathcal{Q}| \cdot |\mathcal{D}_d|)$ similarity computations. By restricting re-ranking to a candidate set $\mathcal{R}(\mathcal{Q})$ of size $N$, total complexity becomes $O(N \cdot |\mathcal{Q}| \cdot \overline{|\mathcal{D}_d|})$ where $\overline{|\mathcal{D}_d|}$ denotes the average number of child embeddings per parent in the candidate set.

Stage 1 is designed to preserve high recall of relevant parents while reducing computational cost, with Exact MaxSim ensuring final ranking fidelity.

\subsection{Indexing}

During indexing, documents and media are segmented into retrieval units appropriate to each modality, such as pages, images, or temporal video segments. Each unit is encoded as a set of dense embeddings in a shared multimodal space: text as token embeddings, images as patch embeddings, and video as frame or region level embeddings. Each retrievable unit is stored as a parent document, with its embeddings stored as child documents containing a single dense vector field for ANN search. Parent and child documents may also carry metadata used for filtering and scoping. The indexing pipeline is represented visually in Figure~\ref{fig:system_overview}.

\subsection{Retrieval}

\begin{figure*}[t]
\centering
\includegraphics[width=0.9\textwidth]{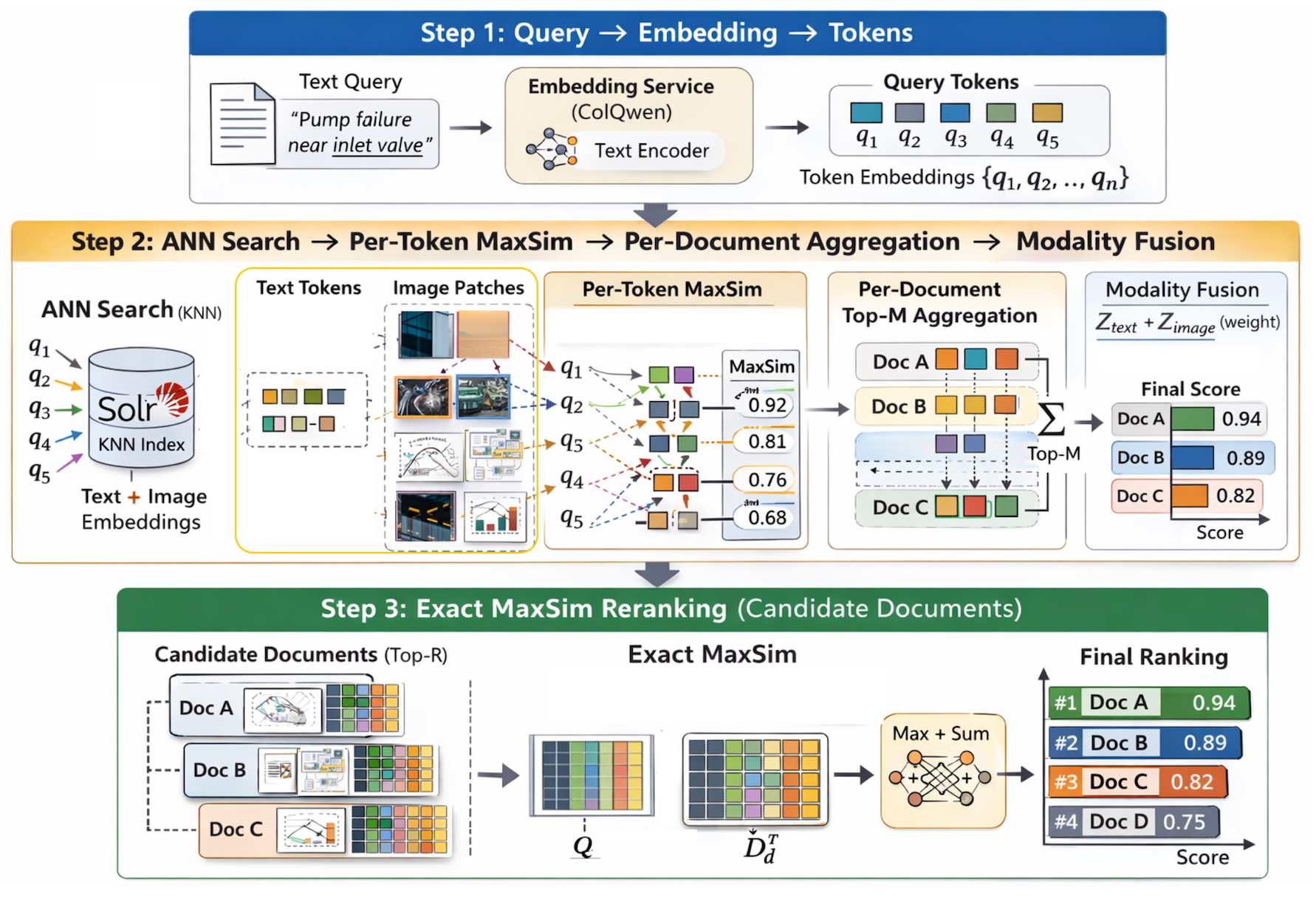}
\caption{Retrieval Pipeline showing the query embedding, the approximate candidate generation and the exact MaxSim ranking.}
\label{fig:retrieval_pipeline}
\end{figure*}

Figure~\ref{fig:retrieval_pipeline} illustrates the complete retrieval flow.

\subsubsection{Query Embedding}

At query time, a textual query is encoded into token level embeddings using the same encoder used for indexing.

\subsubsection{Stage 1: Parallel Token Level Candidate Generation}

The first stage of AMES performs an approximate candidate generation using ANN search over child embeddings. Given a query consisting of $|\mathcal{Q}|$ token vectors, we issue one ANN request per token vector. These requests can be executed asynchronously in parallel with bounded concurrency to mitigate fan-out latency while avoiding excessive backend load. Each ANN request retrieves the top-K most similar child embeddings with \texttt{numCandidates} controlling search breadth. Structured filters can be applied directly at the child-document level to enforce metadata constraints during this retrieval.

When multiple modalities are indexed (e.g., extracted text tokens and image patches), ANN search is executed separately per modality using modality specific filters. This yields multiple approximate candidate sets: for example one derived from text embeddings and one from visual embeddings.

For each parent document $d$ and query token $\mathbf{q}_i$, we retain only the maximum similarity observed across all matching child embeddings. Rather than summing across all query tokens, we apply a per-document top-M aggregation strategy: for each parent, we retain the M highest token level maxima and sum only those values to produce an approximate score. This top-M aggregation reduces the influence of low signal or weakly grounded query tokens while preserving strong localized matches.

Modality specific scores are normalized using robust statistics (median and median absolute deviation) to mitigate scale differences between modalities. Final Stage 1 candidate scores are computed as a weighted combination of normalized scores from each modality, with equal weighting in our default configuration. The top N ranked parent documents are passed to the exact re-ranking stage.

\subsubsection{Stage 2: Exact MaxSim Re-ranking}

In the second stage, AMES re-ranks the top-N candidate parent documents using full Exact MaxSim computation. For each shortlisted parent, all associated child embeddings are retrieved. Query and document embeddings are L2-normalized, enabling similarity computation via dot product.

This computation can be implemented in PyTorch using batched matrix multiplications. By restricting Exact MaxSim to a small candidate set, the two-stage Approximate $\rightarrow$ Exact design achieves high retrieval fidelity while maintaining scalability within a standard enterprise search stack.

\subsubsection{Video Handling}

Video is handled by segmenting each asset into semantically coherent temporal units, which serve as parent documents. Frame or region level embeddings are indexed as child documents in the same shared space used for text and images. Retrieval then follows the same two stage pipeline, with the main practical difference being the larger number of child embeddings per parent.

\subsection{Infrastructure Assumptions}

AMES assumes a backend that supports ANN search, structured filtering, grouping of child embeddings under parent entities, and efficient retrieval of child embeddings for shortlisted parents. These capabilities are available in most modern vector databases and vector-enabled search engines.

\section{Reference Implementation in Solr}

We instantiate AMES in Apache Solr using native KNN search and parent--child documents. Because Solr's vector search operates at the document level and does not natively support token level late interaction, each embedding is indexed as an individual child document and parent level aggregation is performed client side. This design enables multi-vector retrieval while remaining compatible with Solr's standard indexing and filtering mechanisms.

\subsection{Parent--Child Schema in Solr}

Each retrievable unit, such as a document page, image, or video segment, is represented as a parent document. Parent documents store metadata used for identification, filtering, and scoping, but do not participate directly in vector search.

Fine-grained embeddings are stored as child documents associated with a single parent. Each child document corresponds to one text token, image patch, or video frame/region embedding, and contains (i) a dense vector field indexed for KNN search, (ii) an \texttt{embedding\_source} field indicating modality, and (iii) inherited metadata fields required for filtering. All child embeddings share a common dimensionality and are produced in a shared embedding space, enabling retrieval across textual and visual content within the same index.

This design increases index cardinality because the number of indexed documents scales with the number of child embeddings rather than the number of parent units, but it preserves fine-grained retrieval semantics within Solr's standard document model.

\subsection{Vector Indexing and Metadata Filtering}

Approximate nearest neighbor search is performed over child documents using Solr's native KNN support. At query time, each query vector retrieves candidate child embeddings, and structured constraints are incorporated directly into retrieval through Solr filter queries (\texttt{fq}). These filters may enforce project scope, document type, modality, or other metadata constraints during ANN search rather than after retrieval.

When multiple modalities are indexed under a parent, modality-specific retrieval is enabled by filtering on \texttt{embedding\_source}. This allows textual and visual ANN retrieval to be executed separately while preserving a unified parent--child representation.

\subsection{Parent-Level Aggregation Semantics}

Late interaction scoring requires aggregating token-level similarities at the parent document level. Solr does not natively implement MaxSim aggregation. AMES computes token-level maxima and parent-level aggregation externally after ANN retrieval. Because ANN search operates over child embeddings, aggregation is performed client-side after retrieval. For each parent document $d$ and query token $\mathbf{q}_i$, only the maximum similarity over all matching child embeddings under that parent is retained. This enforces the standard MaxSim semantics at the token level:
\[
s_i(d) = \max_{j \in d} \langle \mathbf{q}_i, \mathbf{d}_j \rangle
\]

Stage 1 candidate generation applies a per-document top-M aggregation over these token level maxima to produce approximate scores. Stage 2 re-ranking recomputes the full MaxSim score over all child embeddings for shortlisted parents. This separation of ANN search (child-level) and aggregation (parent-level) preserves the theoretical formulation of late interaction retrieval while leveraging Solr's existing indexing and filtering infrastructure.

AMES implements exact re-ranking using PyTorch, executed on accelerator hardware when available. We support execution on GPUs via CUDA and Apple Silicon via Metal Performance Shaders (MPS). To improve efficiency, computations are performed using fp16 precision on supported devices, with normalization applied directly on the device to minimize data transfer and synchronization overhead.

\subsection{Enterprise Compatibility}

Because AMES is implemented within Solr's native document model, it preserves structured filtering, compatibility with existing metadata and access-control pipelines, shard-based scalability, and coexistence with lexical fields and hybrid retrieval workflows. No external vector database or specialized ANN service is required.

\section{Retrieval Method}

AMES instantiates the two-stage retrieval objective defined in Section 3 using an approximate candidate generation stage followed by exact reranking. We follow the formal definitions introduced in Section 3.

\subsection{Stage 1: Parallel Token Level ANN Candidate Generation}

For each query token $\mathbf{q}_i$, we issue an approximate nearest neighbor (ANN) search over child embeddings. These searches are executed asynchronously in parallel with bounded concurrency. Each ANN query retrieves the top-K most similar child embeddings using dot product similarity. Structured constraints are applied at the child-document level via filter queries.

When multiple modalities are indexed (e.g., text tokens and image patches), ANN search is executed separately per modality using modality-specific filters. This yields modality-specific candidate sets.

\subsubsection{Per Token MaxSim Approximation}

For a parent document $d$, we retain only the maximum similarity observed for each query token:
\[
s_i(d) = \max_{j \in d} \langle \mathbf{q}_i, \mathbf{d}_j \rangle
\]

This enforces standard MaxSim semantics at the token level while restricting computation to ANN-retrieved child embeddings.

\subsubsection{Per-Document Top-M Aggregation}

Summing over all query tokens can introduce noise from weakly grounded or low-signal tokens. To mitigate this effect during candidate generation, we apply a per-document top-M aggregation strategy.

For each parent document $d$, let
\[
S(d) = \{s_i(d)\}_{i=1}^{|\mathcal{Q}|}
\]
denote the set of token-level maxima. Let $\text{TopM}(S(d))$ be the M largest elements of this set.

The approximate Stage 1 score is then:
\[
\text{ApproxMaxSim}_M(\mathcal{Q}, d) = \sum_{s \in \text{TopM}(S(d))} s
\]

This aggregation retains only the strongest token level alignments for each document, reducing score dilution while preserving recall during candidate generation. When multiple modalities are indexed under a parent document, the Top-M aggregation described above is applied independently within each modality. That is, we compute a modality specific approximate score for text embeddings and a separate approximate score for visual embeddings. These modality specific scores are then combined through a fusion step described below.

\subsubsection{Modality-Aware Fusion}

If there are multiple modalities in the corpus then they need to be treated separately. For example, let $A_{\text{text}}(d)$ and $A_{\text{image}}(d)$ denote the modality-specific Top-M approximate scores computed for parent document $d$. The score distributions may differ across modalities therefore we apply robust normalization before combining them. When both textual and visual embeddings are present, Stage 1 is executed independently per modality. The normalization is achieved using median and median absolute deviation (MAD):
\[
Z_m(d) = \frac{A_m(d) - \text{median}(A_m)}{\text{MAD}(A_m)}
\]
where $m \in \{\text{text}, \text{image}\}$

Final Stage 1 candidate scores are computed via weighted fusion:
\[
A_{\text{fused}}(d) = w_{\text{text}} Z_{\text{text}}(d) + w_{\text{image}} Z_{\text{image}}(d)
\]

In our experiments, we use equal weights:
\[
w_{\text{text}} = w_{\text{image}} = 0.5
\]

This approach can be used across any modalities.

\subsection{Stage 2: Exact MaxSim Reranking}

In the second stage, AMES re-ranks a shortlist of the top-N candidate parent documents using exact MaxSim. For each shortlisted parent, all associated child embeddings are fetched, and relevance is recomputed. Unlike Stage 1, no top-M truncation is applied; all query tokens contribute to the final score.

Similarity matrices are computed using batched matrix multiplication:
\[
S = \mathcal{Q} \mathcal{D}_d^\top
\]
followed by a max reduction over document embeddings and a sum over query tokens. Mixed precision (fp16) execution is used on supported devices to improve throughput while maintaining numerical stability.

By restricting Exact MaxSim to a small candidate set $\mathcal{R}$, AMES achieves full late-interaction fidelity without incurring prohibitive computational cost.

\section{Experiments}

We evaluate AMES on a large scale multimodal benchmark under realistic retrieval settings to assess whether a Solr native Approximate $\rightarrow$ Exact MaxSim pipeline can achieve competitive ranking quality while remaining compatible with enterprise search infrastructure. Our focus is on retrieval effectiveness in a production style two-stage pipeline rather than isolated embedding quality.

\subsection{Experimental Setup}

We evaluate on the ViDoRe V3 benchmark~\cite{vidore2026}, which contains visually rich, multi-page enterprise-] style documents paired with human annotated queries and graded relevance judgments. Following the official protocol, we perform page level retrieval, ranking document pages independently for each query.

To reflect our target deployment setting, we restrict evaluation to the English document, English query Industrial subset of ViDoRe V3. In all experiments, queries are textual, so retrieval measures text-to-multimodal grounding against pages containing extracted text and visual content.

We implement dense pooled, and token-level late-interaction retrieval methods within Apache Solr and report ranking quality using nDCG@K, consistent with the ViDoRe benchmark. To contextualize our results, we compare against the English-only results reported in Table~9 of ViDoRe V3, which includes several strong late-interaction and multi-vector baselines.

Our Solr based AMES configuration uses parallel ANN-based candidate generation followed by Exact MaxSim reranking over the retrieved candidates. Unless otherwise stated, we use per-token ANN retrieval with top-$K=10$, candidate pool size $N=80$, Top-$M$ aggregation with $M=12$, and \texttt{numCandidates}$=250$. These hyperparameters are held fixed across embedding backbones.

\subsection{Results}

Table~\ref{tab:results} reports English-only retrieval performance on the ViDoRe V3 Industrial dataset. We evaluate AMES across multiple embedding backbones and compare against the current SOTA nDCG@10 results reported in the ViDoRe technical report and public leaderboard.

Across all evaluated backbones, our Approximate MaxSim implementation achieves competitive performance relative to previously reported English-only SOTA results. Notably,

\begin{itemize}
\item No modifications are made to the embedding models.
\item Improvements arise solely from retrieval strategy.
\item Gains are achieved under a scalable two-stage architecture implemented entirely within Solr.
\end{itemize}

These results indicate that AMES achieves a favorable balance between retrieval scalability and ranking precision, enabling competitive late-interaction performance within a production-ready Solr-based system. Importantly, these results are obtained without modifying the embedding models or introducing specialized ANN infrastructure.

\begin{table*}[!t]
\centering
\small
\caption{English-only retrieval performance on ViDoRe V3 Industrial Dataset (NDCG)}
\label{tab:results}
\begin{tabular}{lccccc}
\toprule
& \multicolumn{4}{c}{\textbf{AMES}} & \textbf{SOTA} \\
\cmidrule(lr){2-5}
\textbf{Model} & \textbf{@1} & \textbf{@3} & \textbf{@5} & \textbf{@10} & \textbf{@10} \\
\midrule
ColQwen3.0 & 62.5 & 56.6 & 57.2 & \textbf{58.1} & 54.1 \\
ColQwen2.5 & 53.7 & 49.9 & 51.0 & \textbf{52.4} & 49.4 \\
ColQwen2.5-Omni & 53.7 & 50.3 & 50.4 & \textbf{52.2} & 49.4 \\
\bottomrule
\end{tabular}
\end{table*}

\subsection{Limitations and Future Work}

Our evaluation is limited to the English-only Industrial subset of ViDoRe V3 and does not yet include a systematic ablation of Stage~1 recall, Top-$M$ aggregation, candidate set size, or Exact MaxSim reranking. In addition, while AMES is designed for scalable enterprise deployment, we do not report a full systems analysis of latency, throughput, or memory usage. These remain important directions for future evaluation.

\section{Conclusion}

In this work, we presented AMES, a backend agnostic architecture for multimodal late-interaction retrieval that integrates fine grained multi-vector representations within standard enterprise search infrastructure. By combining parallel token level ANN candidate generation with Exact MaxSim re-ranking, we demonstrate that late interaction retrieval can be realized without specialized retrieval engines or external vector databases. The AMES reference implementation in Apache Solr shows that parent/child indexing, structured filtering, and native vector search capabilities are sufficient to support token, patch, and frame-level retrieval across text, image, and video modalities within a unified pipeline. Experiments on the ViDoRe V3 benchmark indicate that this Approximate $\rightarrow$ Exact strategy achieves competitive ranking performance relative to prior late-interaction systems, despite operating within a general purpose search engine. More broadly, these findings indicate that AMES demonstrates how modern multi-vector late-interaction methods can be integrated into production search environments using existing infrastructure capabilities, reducing the disconnect between research oriented retrieval models and enterprise deployment constraints.

\end{document}